# Machine Learning Assisted Revelation of the Best Performing Single Hetero-junction Thermophotovoltaic Cell

Ahnaf Tahmid Abir, Arifuzzaman Joy, and Jaker Hossain[*]

*Solar Energy Laboratory, Department of Electrical and Electronic Engineering, University of Rajshahi, Rajshahi 6205, Bangladesh.*

**Abstract**

In this work, Machine Learning (ML) techniques have been employed to explore the highest performing single-heteronunction thermophotovoltaic cell. Initially, traditional homo junction TPV cells have been explored using ML methodologies for the optimal material combinations. ML methods have notably been devoted to analyze the importance of each parameter in the model, thereby improving the comprehension of the system's behavior and facilitating design optimization. Following this investigation, it has been found that Ge emerged as the most effective emitter layer when paired with the optimal base layer, InGaAsSb compound that possesses a direct bandgap of 0.53 eV. Subsequently, a p-Ge/n-InGaAsSb single-heterojunction TPV cell is introduced executing a device transport model featuring a p-n structure. This cell operated at black body ($T_{BB}$) and cell temperatures of 1578 K and 300 K, respectively. Through meticulous optimization efforts, the performance of the TPV cell is significantly enhanced resulting in an impressive efficiency of 16.50%. This efficiency is accompanied by a short circuit current, $J_{SC}$=15.53 A/cm$^2$, an open-circuit voltage, $V_{OC}$=0.47 V, and a fill factor FF=79.5%. These findings suggest that this structural configuration holds considerable promise for the development of high-performance TPV cells.

## 1. Introduction

Thermophotovoltaic (TPV) cell generally a semiconductor having a structure of p-n or n-p gives electrical energy by utilizing the heat energy [1]. Actually, this kinds of cells have introduced a way to exploit the waste heat, in this sense its analogous to a blessing for this energy thrust world. That's the reason, for the TPV cells being an ancient member invented in 1960 and first applied in the space application in 1970 [2]. The 1980 is the year when a renowned researcher named Swanson has found the potential of silicon as a TPV cell, associated with an emitter consists of tungsten, provided 29% efficiency at an operating temperature of 2000 °C [3]. After that, this field has observed a satisfactory success with an efficiency up-to 40% at 2400 K and interestingly, it can provide 27% at 600 K which indicates a great improvement [2]. A prediction of a tremendous efficiency up-to 56% at 2500 K is made by the researchers which can make the TPV cells as popular as solar cells [2]. A TPV system generally consists of a heat source, an emitter and a TPV cell. In this system, the role of emitter is to radiate photons towards the TPV cell having low energy gap [4]. The ideal bandgap requirement for a TPV cell is ranging between 0.5 eV and 0.7 eV so that it can operate in near the IR(Infrared) range with the thermal radiation intensity from 1000 K to 2000 K [1,5]. The conventional materials for the TPV cells are indium gallium arsenide antimonide (InGaAsSb), indium gallium arsenide (InGaAs), indium antimonide (InAs), lead selenide (PbSe), germanium (Ge), geranium tin (GeSn), gallium antimonide (GaSb) etc. The margin of the efficiency of those traditional single homo-junction TPV cell fall between 1% and 11% [5]. A hetero-junction can boost the efficiency limit, as it serves better passivation quality than a homo-junction [6]. However, find out the best combination between the emitter and the base to achieve the best efficiency is a tedious work for human being if they try it manually. ML can be appointed to minimize human effort in selecting the optimal combination of different materials.

Machine learning belongs to the group of artificial intelligence (AI), which explores the computer algorithms that enable systems to learn and enhance their performance based on experience just like a child. These algorithms empower systems to autonomously make decisions by uncovering significant patterns within intricate datasets, thereby eliminating the need for external assistance [7]. Likewise, rather than encoding information directly into computers, ML aims to acquire meaningful insights and patterns from examples and observations automatically [8]. Progress in ML has facilitated the emergence of intelligent systems in recent times, possessing cognitive abilities akin to those of humans, which are increasingly integrated into our professional and personal spheres [8-9]. Conventionally, ML has been linked with the manipulation and analysis of signals and images, evoking thoughts of applications such as autonomous vehicles, processing the natural language, and the recognition of the optical characters. Presently, numerous industrial sectors are dedicating considerable resources to integrating ML methods [10]. Furthermore, ML can aid in handling the vast amounts of data generated by empirical results and calculations. Deriving significant scientific insights from those data can be challenging and time-consuming, often comparable to the effort required for the experiments or computations themselves [10]. Hence, adopting effective data preprocessing techniques is essential for the successful advancement of any ML applications. The subsequent step involves selecting an appropriate machine learning model which falls into four main categories: Guided, partially guided, unguided, and reward-based learning. Guided models utilize labeled datasets during training to get the expected outcomes. Conversely, unguided models are trained on unlabeled datasets without guidance. Partially guided models combine elements of both supervised and unsupervised learning, using a small set of labeled samples along with a larger set of unlabeled samples. Reinforcement learning models, on the other hand, learn through a system of rewards and punishments based on desired and undesired actions [11]. Gradient boosting stands as a formidable machine-learning method renowned for delivering cutting-edge outcomes across diverse applications. It involves assembling weak predictors and iteratively refining them through gradient descent in a functional realm to create a robust predictor. This technique has garnered immense success across an extensive spectrum of machine learning endeavors [12]. After investigating the traditional TPV cells it predicts Ge as the best emitter and InGaAsSb as the best base.

The potential of InGaAsSb as a TPV base has been proven by various studies as it can comfortably response between 1500 nm to 2500 nm wavelength [13]. Another advantage of using InGaAsSb is, this kinds of diodes perfectly lattice matched with the substrate provides a huge probabilities in the band-gap engineering term [14]. Moreover, the most attractiveness of this material as a TPV base is its satisfactory transport of minority charge carriers [14]. The observed minority carrier lifespan in n-type dual hetero-structure (DH) InGaAsSb samples indicate a value of approximately 1 microsecond and carrier surface velocities ranging from approximately 700 to 2000 cm/s at the interfaces [15]. It is recognized that the use of tellurium (Te) for doping to achieve n-type InGaAsSb layers having the concentration almost $4 \times 10^{18}$ cm$^{-3}$ significantly influences the quality of the interface between the heterojunction epitaxial layer and the substrate. This is due to the high diffusion coefficient of the Te atoms [16-17]. Liquid phase epitaxy (LPE) method is popular for the production of this material [13]. The growth solutions consist of high-purity metals such as indium, gallium, and antimony, along with undoped arsenic. After baking at 700°C for fifteen hours under flowing hydrogen to remove impurities, dopant impurities are added before growth [13].

Ge is suitable for use as the emitter in TPV cells as it exhibits a lower absorption coefficient due to its status as an indirect band-gap material. Despite this, Ge is readily available and offers a cost-effective option for manufacturing TPV cells [18]. Additionally, Ge cells demonstrate a higher temperature coefficient in comparison to GaSb or InGaAsSb, further supporting Ge's efficacy as an optimal emitter or window layer due to its proximity to the heat source [19]. To create a p-type Ge emitter in an n-InGaAsSb, p-type dopants such as boron (B), aluminium (Al), gallium (Ga), and indium (In) are typically employed. The spread of p-type dopants mirrors the process of self-diffusion, and in the instance of B, it's even considerably slower. The gradual dispersion of the acceptor dopants, especially B, is highly beneficial for creating ultra-thin regions of acceptor-doped Ge [20]. In contrast to Boron (B), the diffusion of other p-type dopants aligns completely with the vacancy mechanism. The marginally elevated diffusion activation enthalpy of Al, Ga, and In compared to self-diffusion indicates an interaction between these p-type dopants and vacancies that is less appealing than with n-type dopants such as phosphorus (P), arsenic (As), and antimony (Sb). This variation in diffusion behavior between p- and n-type dopants in Germanium (Ge) is likely attributable to coulomb interactions between the substitutional dopant and the vacancy [20].

In this study ML is applied to find out the proper combination of the emitter and base by analyzing the data of the traditional TPV cells. The prediction of ML is Ge as the emitter and InGaAsSb as the base. Subsequently, the performance of the predicted combination is investigated thoroughly by using the transport model. The approach of this study may have committed as the guideline for the researchers to invent more efficient TPV cells in the future.

## 2.1 Simulating device transport in TPV cells and conducting computations

The equation linking density of current (J) and voltage (V) in a p-n illuminated junction diode stems from traditional PV cell principles [21] and is formulated as follows:

$$J = J_{SC} - J_0 \left[ \exp\left(\frac{qV}{K_B T}\right) - 1 \right] \quad (1)$$

In this scenario, $J_{SC}$, $J_0$, along with $K_B$, T and q stand for the density of current at zero load, the density of current in the absence of light, the constant of Boltzmann, the temperature of the device, and electron charge, accordingly. The determination of $J_{SC}$ can be accomplished utilizing the subsequent formula [5],

$$J_{SC} = \int_0^{\lambda_m} q\phi(\lambda) \text{IQE}(\lambda) d\lambda \quad (2)$$

In the given scenario, $\phi(\lambda)$ denotes the incoming light spectrum of irradiation, and IQE($\lambda$) represents the internal quantum yield. The calculation of $J_{SC}$ entails integrating the product of q, $\phi(\lambda)$, and IQE from 0 to the cutoff wavelength, $\lambda_m$ [1].

The collective IQE of a simple p-n junction comprises the individual IQEs of the emitter, absorber, and depletion region [1,22].

The intrinsic doping density $n_i$ can be calculated using the equation below [1]:

$$n_i = \sqrt{N_c N_v} e^{\frac{-E_g}{2K_B T}} \tag{3}$$

In the formula, $E_g$ represents the semiconductor's energy gap, while $N_C$ and $N_V$ denote the effective density of states within the conduction and valence bands, respectively. These values can be computed using the presented equations [1]:

$$N_c = 2\left(\frac{m_e^* K_B T}{2\pi \hbar^2}\right)^{\frac{3}{2}} \tag{4}$$

$$N_v = 2\left(\frac{m_h^* K_B T}{2\pi \hbar^2}\right)^{\frac{3}{2}} \tag{5}$$

In the expression, $m_e^*$ and, $m_h^*$ represent the actual electrons and holes masses, accordingly. It's crucial to highlight that the IQE is greatly impacted by the coefficient of the absorption, which can be calculated as follows [1,5]:

$$\alpha(\lambda) = \frac{2\pi \lambda q^2 (2m_r)^{1.5}}{n\varepsilon_0 (cm_0)^2 h^3} \left(h\nu - E_g\right)^{0.5} \langle |P_{cv}|^2 \rangle \tag{6}$$

The $m_r$ is for the transformed mass, and $P_{cv}$ is for the matrix of optical transitions.

The formula used to calculate the current density in the absence of light is as follows [5],

$$J_0 = \frac{qn_i^2 D_e}{N_a L_e} \left\{ \frac{Sinh\left(\frac{W_E}{L_e}\right) + \frac{S_F L_e}{D_e} Cosh\left(\frac{W_E}{L_e}\right)}{Cosh\left(\frac{W_E}{L_e}\right) + \frac{S_F L_e}{D_e} Sinh\left(\frac{W_E}{L_e}\right)} \right\} + \frac{qn_i^2 D_h}{N_d L_h} \left\{ \frac{Sinh\left(\frac{W_B}{L_h}\right) + \frac{S_B L_h}{D_h} Cosh\left(\frac{W_B}{L_h}\right)}{Cosh\left(\frac{W_B}{L_h}\right) + \frac{S_B L_h}{D_h} Sinh\left(\frac{W_B}{L_h}\right)} \right\} \tag{7}$$

In that formula, $D_e$ and $D_h$ are for the diffusion coefficients of electrons and holes, whereas $\tau_e$ and $\tau_h$ represent their lifetimes, respectively. The formula for $V_{OC}$, FF and efficiency can used from the previous work on TPV [1]. The transport equations have been solved by MATLAB programming language.

**Table 1:** Physical variables utilized in computing the performance of PV of the proposed TPV cell.

| Parameters | InGaAsSb | Ge | Reference |
|---|---|---|---|
| Thickness | 3000 nm | 100 nm | Optimized |
| Energy gap, $E_g$ | 0.53 eV | 0.66 eV | [15] |

| Intrinsic doping level, $n_i$ | $4.5\times10^{13}$ (cm$^{-3}$) | $2.33\times10^{13}$ (cm$^{-3}$) | [15,23] |
|---|---|---|---|
| Doping concentration | $1\times10^{18}$ cm$^{-3}$ | $1\times10^{16}$ cm$^{-3}$ | Optimized |
| Radiative recombination rate constant, $B_{opt}$ | $6.4\times10^{-14}$ (cm$^3$.s$^{-1}$) | $6.4\times10^{-14}$ (cm$^3$.s$^{-1}$) | [5] |
| Auger recombination rate constant, A | $10^{-31}$ (cm$^6$.s$^{-1}$) | $10^{-31}$ (cm$^6$.s$^{-1}$) | [5] |
| Electron effective mass, $m_e^*$ | $0.037m_0$ | $0.22m_0$ | [24] |
| Hole effective mass, $m_h^*$ | $0.32m_0$ | $0.34m_0$ | [25] |
| energy of spin-orbit splitting, $\Delta$ | 0.35 eV | 0.35 eV | [5] |
| Rate of recombination at the front surface, $S_F$ | $10^5$ cm.s$^{-1}$ | $10^5$ cm.s$^{-1}$ | [5] |
| Rate of recombination at the back surface, $S_B$ | $10^5$ cm.s$^{-1}$ | $10^5$ cm.s$^{-1}$ | [5] |
| Index of refraction, n | 3.45 | 4 | [15] |
| Permittivity, | 13 | 16 | Calculated |
| Electron drift velocity, $\mu_e$ | 6800 (cm$^2$.v$^{-1}$.s$^{-1}$) | 3895 (cm$^2$.v$^{-1}$.s$^{-1}$) | [15,23] |
| Hole drift velocity $\mu_h$ | 565 (cm$^2$.v$^{-1}$.s$^{-1}$) | 2500 (cm$^2$.v$^{-1}$.s$^{-1}$) | [15] |

**2.2 Machine learning analysis of variable importance in TPV cell design**

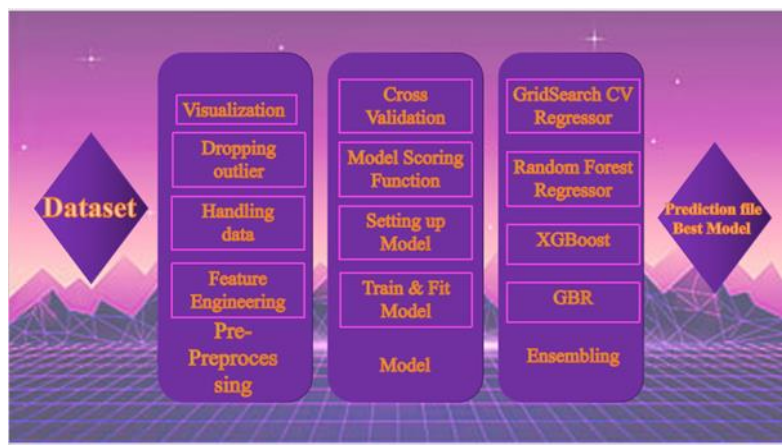

**Figure 1:** The work flow diagram of ML.

Figure 1 presents a visually captivating representation of a machine learning workflow against a backdrop of rich purple hues, evoking curiosity and depth. Divided into distinct sections, it guides viewers through the sequential stages of data processing and model training. The initial phase focuses on data-set handling, encompassing tasks like data cleaning and feature extraction. Moving forward, the model training phase follows, utilizing processed data to train predictive models. Finally, the application of the trained model to new data is depicted, resulting in predictions. This imagery simplifies the intricacies of machine learning, offering an accessible educational resource for understanding its fundamental processes.

The cell's performance is assessed by python programming language using gradient boosting model (GBM) aim to construct predictive models using iterative refinements and non-parametric regressions. Rather than creating a single model, GBM begins by creating a preliminary model and continuously fits noble models through minimizing the function of loss to generate the most accurate model possible. The primary objective of the GBM is to identify a function $F(x)$ that minimizes its loss function $L(y,F(x))$, where $y$ represents the observed outcomes [26].

$$F^* = {}^{\text{argmin}}_F E_{y,x} L(y, F(x)) \tag{8}$$

A boosted model, by definition, is a blended mix of the foundational learners, with each learner weighted accordingly [26].

$$F(x;\{\beta_m, a_m\}{}^M_1) = \sum_{m=1}^M B_m h(x : a_m) \tag{9}$$

This statement refers to a scenario where $h(x : a_m)$ denotes a fundamental learning model characterized by a parameter $a$.

The optimization procedure can be formulated in the following way [26]:

$$P^* = {}^{\text{argmin}}_P \Phi(P) \Phi(P) = E_{y,x} L(y, F(x; P)) F^*(x) = F(x; P^*) \tag{10}$$

In the mentioned equation, $P^* = \sum_{m=0}^M P_m$ and $P_m$ denotes the steps of boosting.

The conventional homo-junction TPV cells composed of Ge, InGaAsSb, InGaAs, GeTe, GeSn, InAs, InGaSb, GaSb, PbSe and $AuCuSe_4$ are thoroughly analyzed by the GBM to find out the best combination to build up the best efficient hetero-junction TPV cell. With those cells 45 combinations are possible. Manually checking the combinations to get the best one is not only tedious but also time consuming. That's why the duty have been transferred to the ML. However doing so isn't an easy task, as it needs too much data to train the model. There are 14 independent variables have been varied and each one has 100 values leads to 14000 data as the input of the machine. The data has been split into two data types, where the 80% data worked as the train data and the 20% data as the test data. After the training procedure the ML has showed it's caliber by predicting the values accurately or close to the actual values. The prediction of the model for both the test and train values has been presented by the regression plot in the Figure 2.

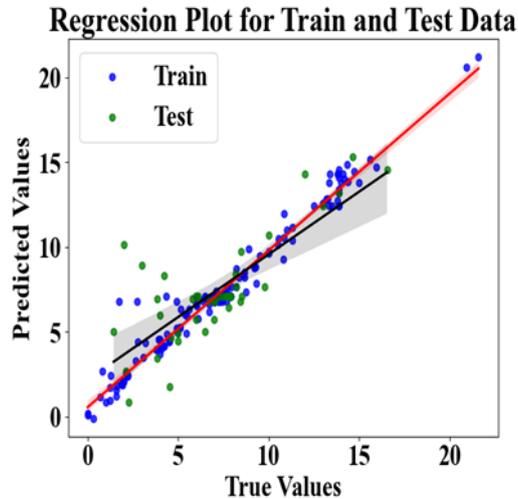

**Figure 2:** Regression plot for train and test data.

In the regression plot, dots are typically color-coded to distinguish between different datasets, with blue color representing the training data and the green representing the test data. While the color of the dots doesn't affect interpretation, it aids in distinguishing between datasets. The curved lines, often called the 'line of best fit' or 'regression line', depict predicted values from the model, with the red line for training data and the black for test data. These lines result from a mathematical process called 'regression', aiming to minimize the distance between data points and the line. Closer alignment between data points and the line indicates more accurate predictions, while wider scattering suggests less accuracy. In machine learning, the goal is to minimize this distance, indicating higher prediction accuracy.

After testing the authenticity of the ML it's time to check the correlation between the inputs and the outputs. The correlation matrix is given in the Figure 3.

The visual representation helps to understand complex data correlations. From Figure 3(a), it is clear that the $J_{SC}$ has obeyed the $T_{BB}$ with a great positivity. Whereas, it shows a hostile behaviour with the energy of spin-orbit splitting, $\Delta$. A moderate positive relation of $J_{SC}$ with the Shockley-Read-Hall (SRH) parameter and n has also be noticed. On the counter end, at Figure 3(b) the $V_{OC}$ has showed a disobedient manner against all the parameters as it happy with its rigid values. The FF also tries to follow the $V_{OC}$ except it has showed a negative relation with the intrinsic carrier, $n_i$ depicted in Figure 3(c). Figure 3(d) allows the efficiency to exhibit it's kinship with the input parameters. The efficiency is positively susceptible with the SRH and the $T_{BB}$, on the contrary a negative susceptibility is recorded with the $n_i$ and the $\Delta$. It also showed a moderate positivity with the $\mu_e$ and negativity with the $m_e$. All of these observations indicate that the base of a TPV cell should be lightly doped. Whereas the emitter should have obtained a good value of $\mu_e$ along with a low value of $m_e$.

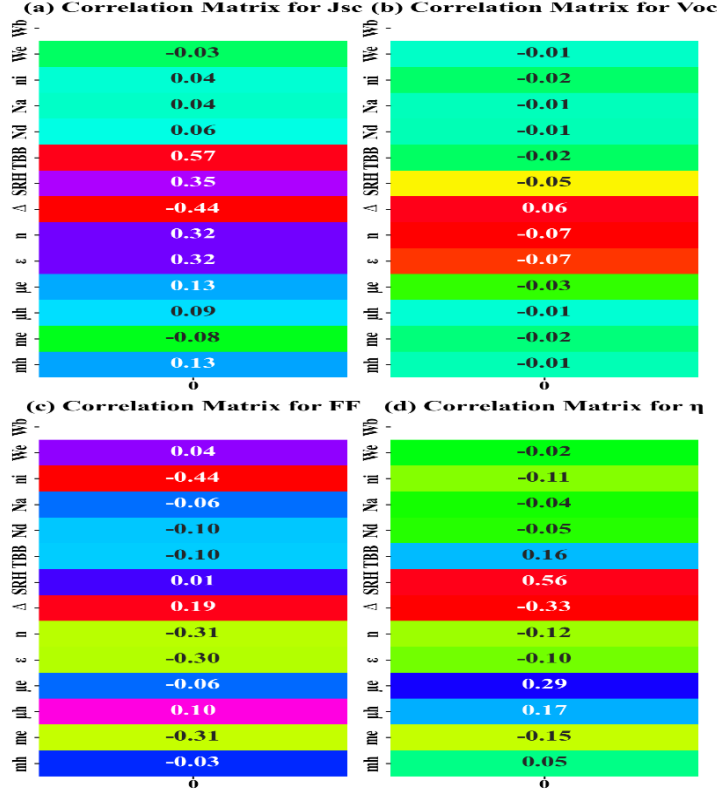

**Figure 3:** Correlation matrix of (a) $J_{SC}$, (b) $V_{OC}$, (c) FF, and (d) η with the input parameters. The image depicts four correlation matrices for $J_{SC}$, $V_{OC}$, FF, and η. These matrices show the relationships between different variables in a TPV cell performance.

In the scenario involving 10 TPV cells, each comprising 14 unique variables namely Base width ($W_b$), emitter width ($W_e$), intrinsic carrier concentration ($n_i$), base doping level ($N_d$), emitter doping level ($N_a$), black body temperature ($T_{BB}$), SRH lifetime (SRH), energy of spin-orbit splitting (Δ), index of refraction (n), permittivity (ϵ), electron mobility ($\mu_e$), hole mobility ($\mu_h$), electron effective mass ($m_e^*$), and hole effective mass ($m_h^*$). The total combinations can be computed by multiplying the options for each cell. With every cell presenting 14 choices, the total combinations amount to $14^{10}$. This results in a vast array of potential combinations, approximately $2.036 \times 10^{12}$. That is why only the parameters which determine the base and emitter are considered and the variables are $n_i$, $\mu_e$ and $\mu_h$. As a result, the total numbers of combinations turn out to be 59049. Along with these, the bandgaps and the electron affinities of the cells are considered to get the precise result. By analyzing all the combinations, the ML takes InGaAsSb as the best base material and Ge as the best emitter or window layer. Some successful combinations with the radiation temperatures and the output variables produced by using the transport model has been presented in the Table 2 which sings the triumph of the model.

**Table 2:** Comparison of various hetero-junction TPV cells

| p-type | n-type | Operating Temperature(K) | $J_{SC}(A/cm^2)$ | $V_{OC}(V)$ | FF(%) | PCE(%) |
|---|---|---|---|---|---|---|
| Ge | InGaAsSb | 1578 | 15.53 | 0.47 | 79.5 | 16.50 |
| InGaAsSb | InGaAs | 1578 | 13.91 | 0.48 | 80.0 | 15.25 |
| InGaAsSb | GeTe | 1578 | 13.00 | 0.41 | 77.5 | 11.92 |
| GeSn | InGaAsSb | 1578 | 14.86 | 0.38 | 76.5 | 12.56 |
| Ge | GeSn | 1550 | 13.87 | 0.35 | 75.0 | 11.29 |
| InGaAs | GeSn | 1550 | 13.15 | 0.35 | 75.0 | 10.65 |
| GeTe | GeSn | 1550 | 12.53 | 0.35 | 75.0 | 10.03 |
| GaSb | GeSn | 1550 | 13.00 | 0.35 | 75.0 | 10.58 |
| Ge | GeTe | 1775 | 21.00 | 0.40 | 77.0 | 11.36 |
| InGaAs | GeTe | 1775 | 20.31 | 0.40 | 77.0 | 10.94 |
| GaSb | GeTe | 1775 | 20.38 | 0.40 | 77.0 | 10.98 |
| Ge | InGaAs | 1950 | 23.55 | 0.54 | 81.5 | 12.62 |
| GaSb | Ge | 1950 | 23.63 | 0.54 | 81.5 | 12.67 |
| InGaAs | InAs | 1035 | 4.00 | 0.30 | 72.0 | 13.97 |

From the table, it is clear that the ML has correctly judge the best combination for a hetero-junction TPV cell. The reason of choosing the GBM model is presented in the next section by comparing it with other two renowned models.

## 3. Results and discussion
### 3.1 Performance comparison of various model to accurately predict the outputs
### 3.1.1 Predictions of random forest regression model

The machine's performance has been scrutinized in details, with four regression plots of Figure 4(a, b, c, d) illustrating the comparison between predicted (blue dots) and actual (red dots) values of various variables (Jsc, Voc, FF, η). These plots depict the effectiveness of a Random

Forest Regressor model in predicting these variables, showcasing both accuracy and errors. Random Forest Regression model belongs to the supervised learning method. Random forests constitute an ensemble learning technique comprised of numerous decision trees, often numbering in the hundreds. By averaging the predictions of each decision tree within the ensemble, random forests mitigate bias and over-fitting, consequently enhancing accuracy [27].

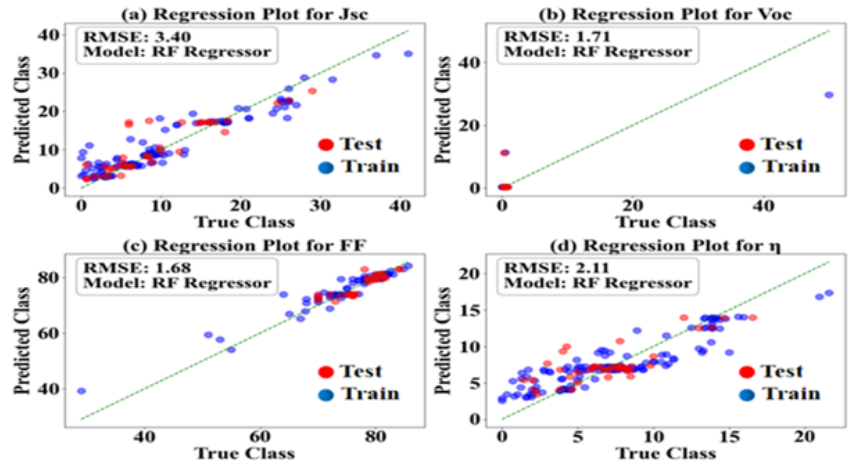

**Figure 4:** Assessing the concurrence between observed and forecasted resultant matrix including (a) $J_{SC}$, (b) $V_{OC}$, (c) FF, and (d) efficiency by the random forest regressor model.

In Figure 4(a), the root means square error (RMSE) is 3.4 for predicting $J_{SC}$. This significant RMSE is attributed to the substantial fluctuation of $J_{SC}$ with changes in $T_{BB}$. Conversely, in Figure 4(b), the ML model achieves near-perfect accuracy when predicting $V_{OC}$, with an RMSE of 1.71. Similarly, the estimation of FF in Figure 4(c) demonstrates good performance, with an RMSE of 1.68. However, in Figure 4(d), the efficiency prediction is more challenging due to the erratic changes in $J_{SC}$, resulting in an RMSE of 2.11 Despite this, the model's overall capability remains unquestioned.

**3.1.2 Predictions of extreme gradient boosting (XGB) regressor model**

Figure 5 illustrates the effectiveness of the XGB regressor model in forecasting output variables, displaying both accuracy and errors. The XGB regressor evaluates the performance of the cell. This powerful machine learning algorithm is well-known for its effectiveness and precision in regression tasks. It operates within the gradient boosting (GB) framework, utilizing weak learners, often decision trees, to create a resilient prediction model. Notably, the XGB regression is adept at managing sizable datasets and provides extensive customization options, enabling the fine-tuning of model performance and the mitigation of over-fitting [28].

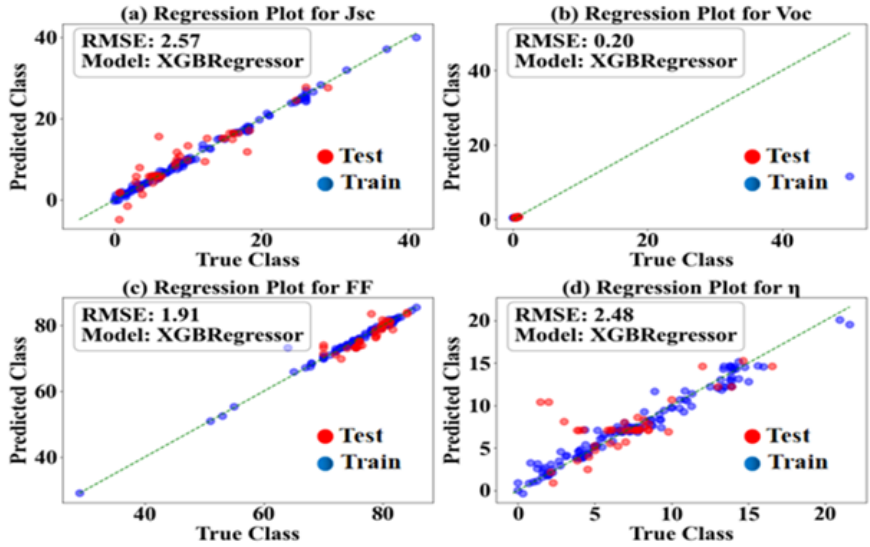

**Figure 5:** Assessing the concurrence between observed and forecasted resultant matrix including (a) $J_{SC}$, (b) $V_{OC}$, (c) FF, and (d) efficiency by XGB Regressor model.

In Figure 5(a), the RMSE stands at 2.57 for $J_{SC}$ prediction. This significant RMSE is due to the substantial $J_{SC}$ fluctuations with $T_{BB}$ changes, as mentioned earlier. Conversely, in Figure 5(b), the ML model achieves nearly flawless accuracy in predicting $V_{OC}$, with an RMSE of only 0.20. Similarly, FF estimation in Figure 5(c) shows commendable performance, with an RMSE of 1.91. However, predicting efficiency in Figure 5(d) proves more challenging due to erratic $J_{SC}$ changes, resulting in an RMSE of 2.48.

### 3.1.3 Predictions of Gradient Boosting (GB) Regression model

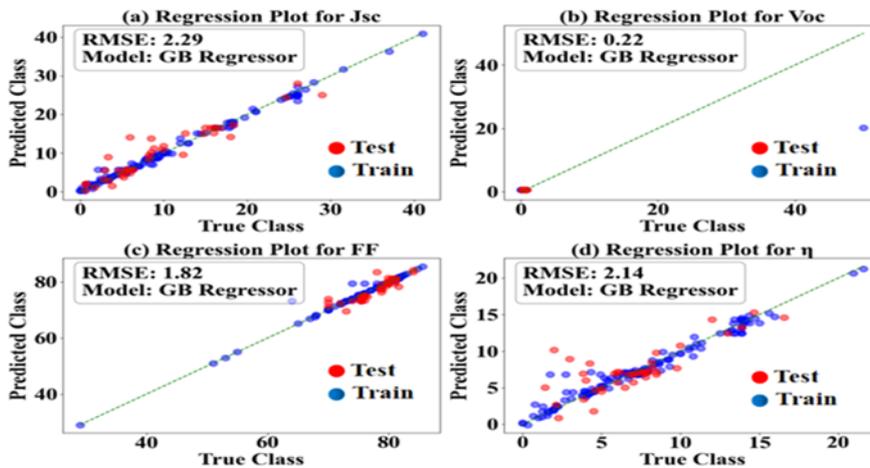

**Figure 6:** Assessing the concurrence between observed and forecasted performance parameters including (a) $J_{SC}$, (b) $V_{OC}$, (c) FF, and (d) efficiency by gradient boosting regressor model.

The plots in Figure 6 showcase the performance of the gradient boosting regressor model in predicting the variables. The visual representation highlights the model's accuracy and errors. Figure 6(a) has depicted that, the root means square error (RMSE) is 2.29 when the ML predict the $J_{SC}$. The abrupt variation of the $J_{SC}$ with the change of $T_{BB}$ causes this considerable number of the RMSE. However, at the time to predict the $V_{OC}$ which is showed in Figure 6(b) the ML has exhibited a perfect accuracy, as the RMSE is only 0.22. The FF also has been estimated nicely with only the RMSE of 1.82 that is presented in the Figure 6(c). The case of efficiency is presented in the Figure 6(d). The abrupt acceleration of the $J_{SC}$, forces the efficiency to change without control, results the value of RMSE 2.14, though this value can't question about the ability of the model. Considering all the resultant parameters, no doubt the GB model has proven itself the superior than the other two models.

### 3.2 Bar chart exploration by using ML

Figure 7 depicts the correlation between output parameters and TPV input parameters by using the GBM model. In Figure 7(a), it is evident that the $J_{SC}$ value experiences significant fluctuations based on the donor doping and the $T_{BB}$ value. Conversely, both $V_{OC}$ and FF are influenced by the base width and intrinsic carrier concentration, as illustrated in Figures 7(b) and 7(c), respectively. Efficiency is influenced by various factors, primarily $n_i$, SRH, $m_e$, and $T_{BB}$ as shown in Figure 7(d). This underscores the importance of careful consideration of $N_a$, $n_i$ and $T_{BB}$ to achieve high efficiency in a TPV cell.

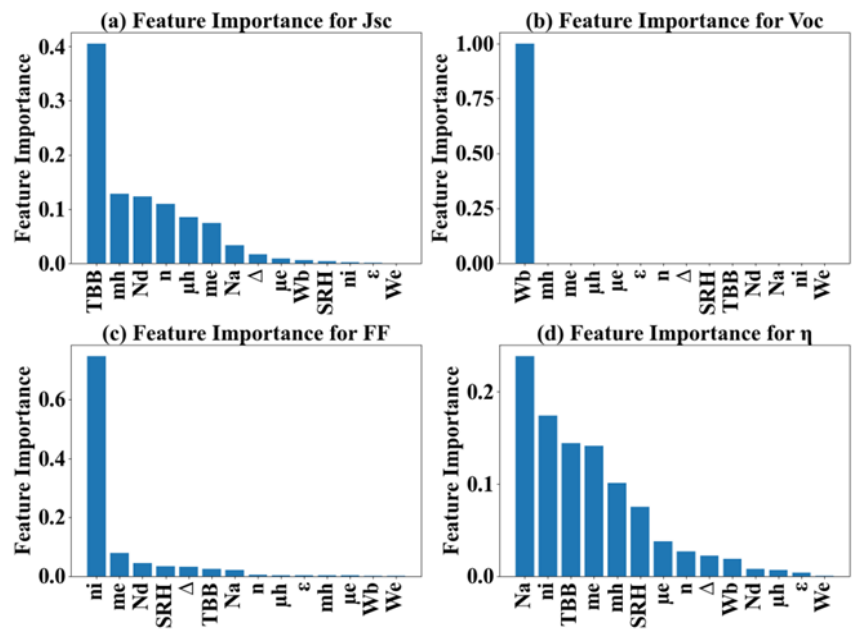

**Figure 7:** The comparative significance of individual features on the outputs.

## 3.3 TPV model structure

Figure 8(a) delimits the short overview of the TPV system. The heart of a TPV system is the heat source. Multiple heat sources are available, with three main types being radioisotope decay, chemical fuel, and concentrated and absorbed sunlight [29]. The heat source elevates the temperature of the emitter to at least 1000 K, leading to the emission of thermal radiation by the emitter. This radiation is then absorbed and converted into electricity by the PV cell [29]. The power density emitted from the TPV emitter is essentially constrained solely by Planck's law for black-body emission. However, the details description of the heat source and the emitter is beyond the scope as this paper only concentrates on the TPV cell. In the presented TPV cell, the role of InGaAsSb is as the base layer whereas the function of Ge is as the window layer. The InGaAsSb material is n-type doped at the concentration of $10^{18}$ cm$^{-3}$ and Ge is p-type doped at the concentration of $10^{16}$ cm$^{-3}$. The optimized thickness for InGaAsSb is 3000 nm whereas that for Ge is 100 nm.

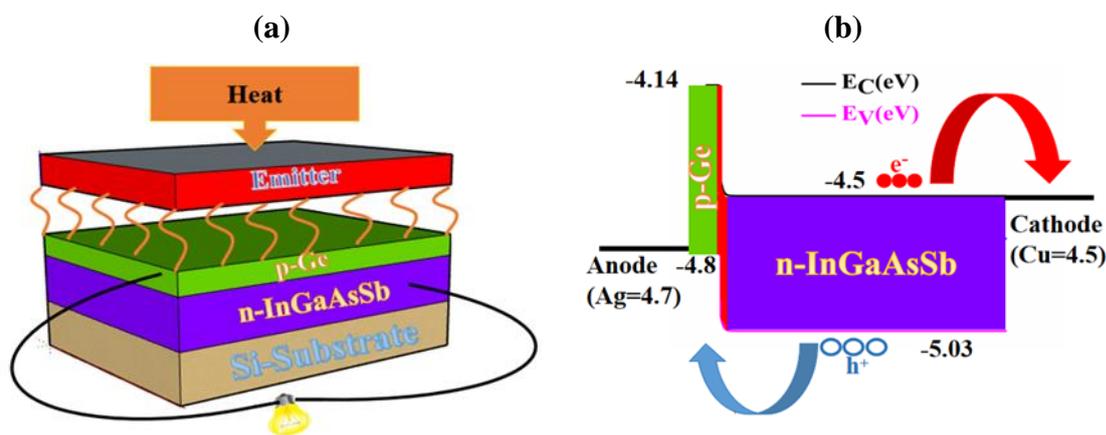

**Figure 8:** (a) A schematic diagram delineating the TPV structure, and (b) an energy band depiction of the TPV cell (not to scale).

Figure 8(b) denotes the energy band structure of the InGaAsSb TPV cell. InGaAsSb with the affinity of electron 4.5 eV [30] and the bandgap of 0.53 eV [15] forms a suitable hetero-junction with Ge having an affinity of electron of 4.14 eV and bandgap of 0.66 eV [25,31]. The hetero-junction formed between the p-type Ge and the n-type InGaAsSb which associates to passivate the carriers towards the electrodes. The trajectories of electrons and holes are depicted by the red and blue-colored arrows accordingly. Using the SCAPS-1D software, which serves as a guiding light in solar cell research, developed by M. Burgelman and colleagues at the University of Gent, Belgium [32], the flat-band potentials for the anode and cathode are estimated to be 4.7 eV and 4.5 eV, respectively, equivalent to the metal work function of Silver (Ag) and Copper (Cu). Hence, metals with work functions ≥4.5 eV and those with work functions ≤4.7 eV can be engaged as appropriate materials for the anode and cathode [1].

### 3.3.1 InGaAsSb base layer

The potential of base layer on the TPV cell is presented in Figure 9 where the variation of thickness is in the x axis and the alteration of doping level is in the y axis. The margin of the alteration of the width and the carrier level are from 1000 nm to 4000 nm and from $10^{15}$ cm$^{-3}$ to $10^{19}$ cm$^{-3}$. In this boundary of the fluctuation of the two independent variables, the $J_{SC}$ is affected greatly, which has been depicted in Figure 9(a). At the doping level $10^{18}$ cm$^{-3}$, the $J_{SC}$ has advanced from about 12 A/cm$^2$ to 17 A/cm$^2$ with the observed range of the base thickness. That means there is a positive relation between the base thickness and the $J_{SC,}$ on the contrary an opposite relationship also examined between the $J_{SC}$ and the doping level of base. The reason behind the amigo behaviour between the thickness and the $J_{SC}$ is the generation of extra carriers at the advanced width [32]. On the other side, the hostility between the $J_{SC}$ and the doping level is the advancement of recombination loss at higher doping level [32].

Figure 9(b) attests the relation of the $V_{OC}$ with the examined two independent variables. The $V_{OC}$ shows almost a constant relation with the width variation, whereas, it is very much susceptible with the impurity density. At the observed impurity density, the $V_{OC}$ fluctuates from approximately 0.3 V to 0.5 V. Actually, the introduction of doping into a semiconductor material disrupts the equilibrium between electrons and holes, resulting in an elevation of the built-in potential across the junction. This effect is commonly noted in semiconductor devices such as diodes and transistors [33-34]. At the case of FF, it follows almost the same trend as the $V_{OC}$, which is depicted in Figure 9(c). This characteristic of the FF can be explained by the by the fact that there exists an empirical relation of FF with $V_{OC}$ that indicates FF increase with the rise in $V_{OC}$ [35].

An interesting thing is noticed in the case of the efficiency as it maintains a positive relation with both the independent variables as depicted in Figure 9(d). The oscillation margin of the efficiency is between 10% and 17%. However, beyond the doping level of $10^{18}$ cm$^{-3}$, it starts to follow a downward path. Mainly at the higher doping, the downward pull of $J_{SC}$ gets domination than the accumulated up ward force of $V_{OC}$ and FF, results the fall of efficiency. Considering those performances and the cost, 3000 nm width and $10^{18}$ cm$^{-3}$ doping value have been optimized.

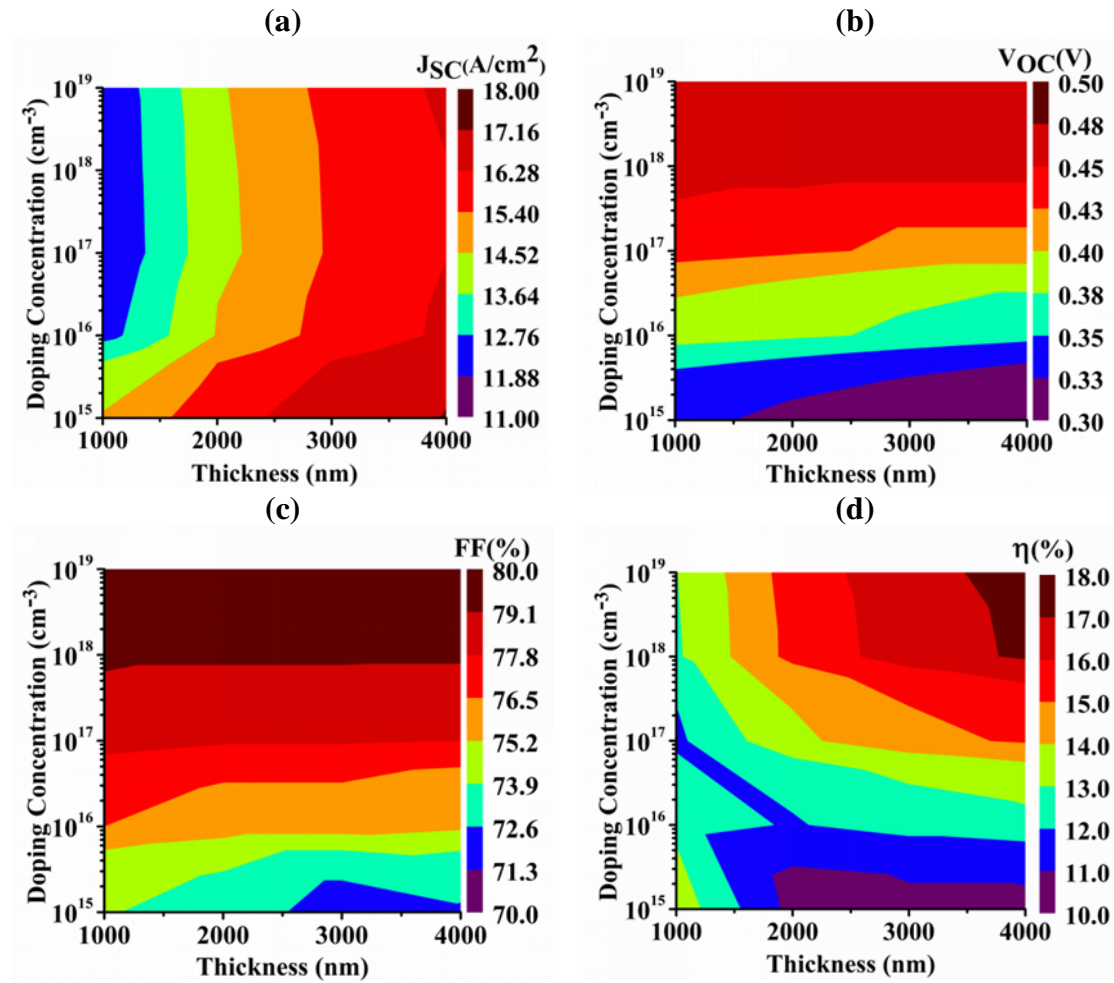

**Figure 9:** The fluctuations in output parameters: (a) $J_{SC}$, (b) $V_{OC}$, (c) FF, and (d) η concerning the width and the impurity (doping) level of the base.

### 3.3.2 Ge Window layer

Figure 10 attests the alterations of the resultant metrics in relation to the fluctuations of the width of the window to the x axis, and the alteration of the impurity concentration of the window to the y axis. The alliance of the $J_{SC}$ with the width and the doping is delineated in the Figure 10(a). The audited range of the width and the carrier are from 100 nm to 500 nm and from $10^{15}$ cm$^{-3}$ to $10^{19}$ cm$^{-3}$. The $J_{SC}$ maintains a negative relation with both the two independent variables, though it is more susceptible with the width than the doping level. At the doping level $10^{16}$ cm$^{-3}$, the $J_{SC}$ commences its travel from around 15.5 A/cm$^2$ at 100 nm and finishes the travel at 12.5 A/cm$^2$. This decrease in current can be linked to the slightly adverse effect of the rising thickness of the window on the IQE along with the advancement of the dark current [1,36]. At 100 nm thickness, the perturbation of doping in the examined range reduces the $J_{SC}$ from about 17 A/cm$^2$ to 14.5 A/cm$^2$. The reduction in minority carrier lifetime caused by the heightened doping concentration has led to a decline in the J$_{SC}$ value [1,36].

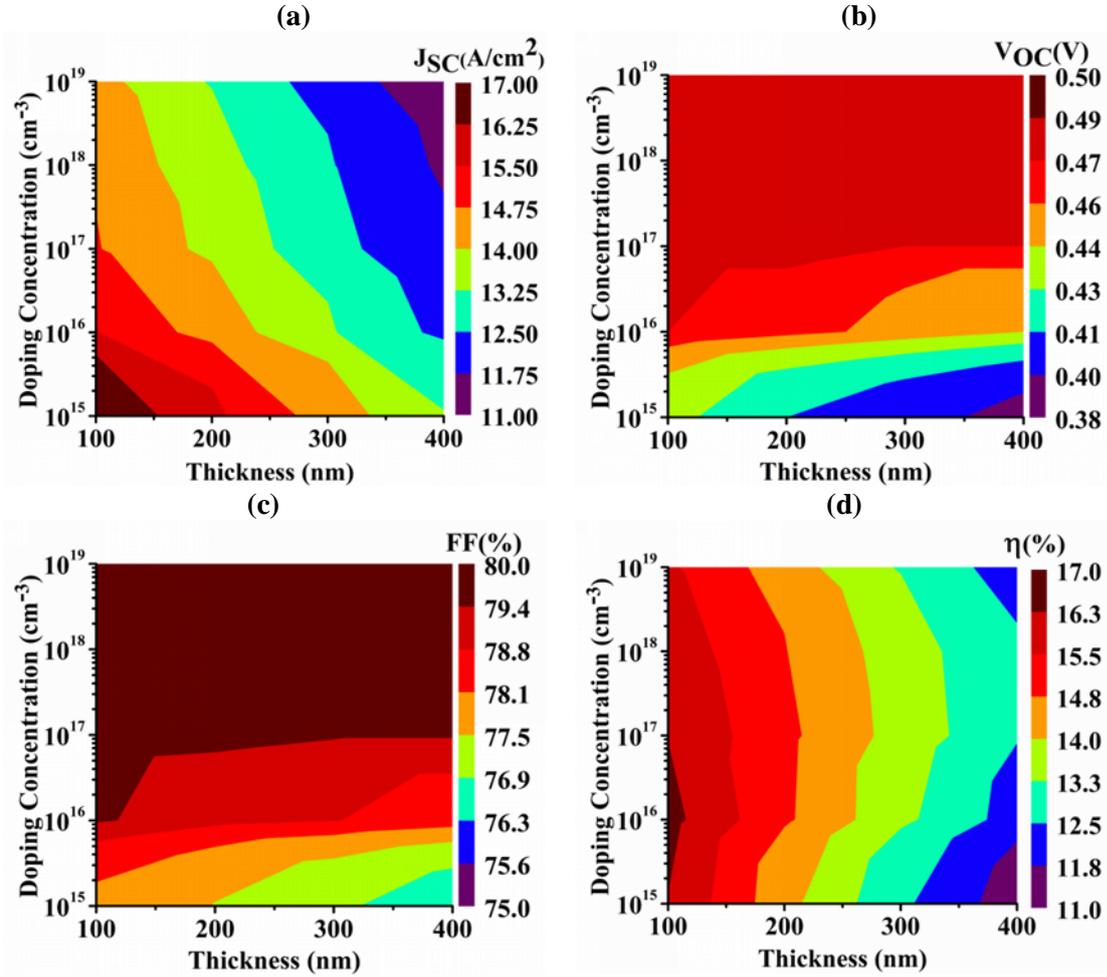

**Figure 10:** The fluctuations in output parameters: (a) J$_{SC}$, (b) V$_{OC}$, (c) FF, (d) η concerning the width and the impurity level of the window.

The oscillation of the V$_{OC}$ with the regarded variables is showed in the Figure 10(b). At 10$^{16}$ cm$^{-3}$, with the switching of the width from 100 nm to 400 nm the V$_{OC}$ falls from 0.47 V to 0.44 V. The V$_{OC}$ declines because of the intricate equilibrium between electrical characteristics and light passage [1]. On the contrary at 100 nm, the V$_{OC}$ initiates to follow an ascensive path up to 10$^{19}$ cm$^{-3}$ and reached the apex of 0.49 V. The characteristics of FF is quite similar to the V$_{OC}$ which is shown in the Figure 10(c) according as FF rise with V$_{OC}$ [35]. All of those three parameters force the efficiency to fall with the enhancement of the width and the doping depicted in Figure 10(d). The peak value of the efficiency is found at 100 nm width and 10$^{16}$ cm$^{-3}$ acceptor density which are optimized for the window layer.

### 3.3.3 Ge- InGaAsSb hetero-junction cell considering various parameters

In Figure 11(a), the fluctuation of photon flux manifests across wavelengths, orchestrated by different $T_{BB}$. As $T_{BB}$ increases, there's a noticeable increase in photon flux, accompanied by a graceful leftward shift of the radiation peak. At the captivating temperature of 1578 K, the TPV cell, akin to a seasoned artist, adeptly harnesses the peak of thermal photons with its cutoff wavelength at 2344 nm. This captivating display reinforces the cell's prominence as a leading figure in the TPV cell domain.

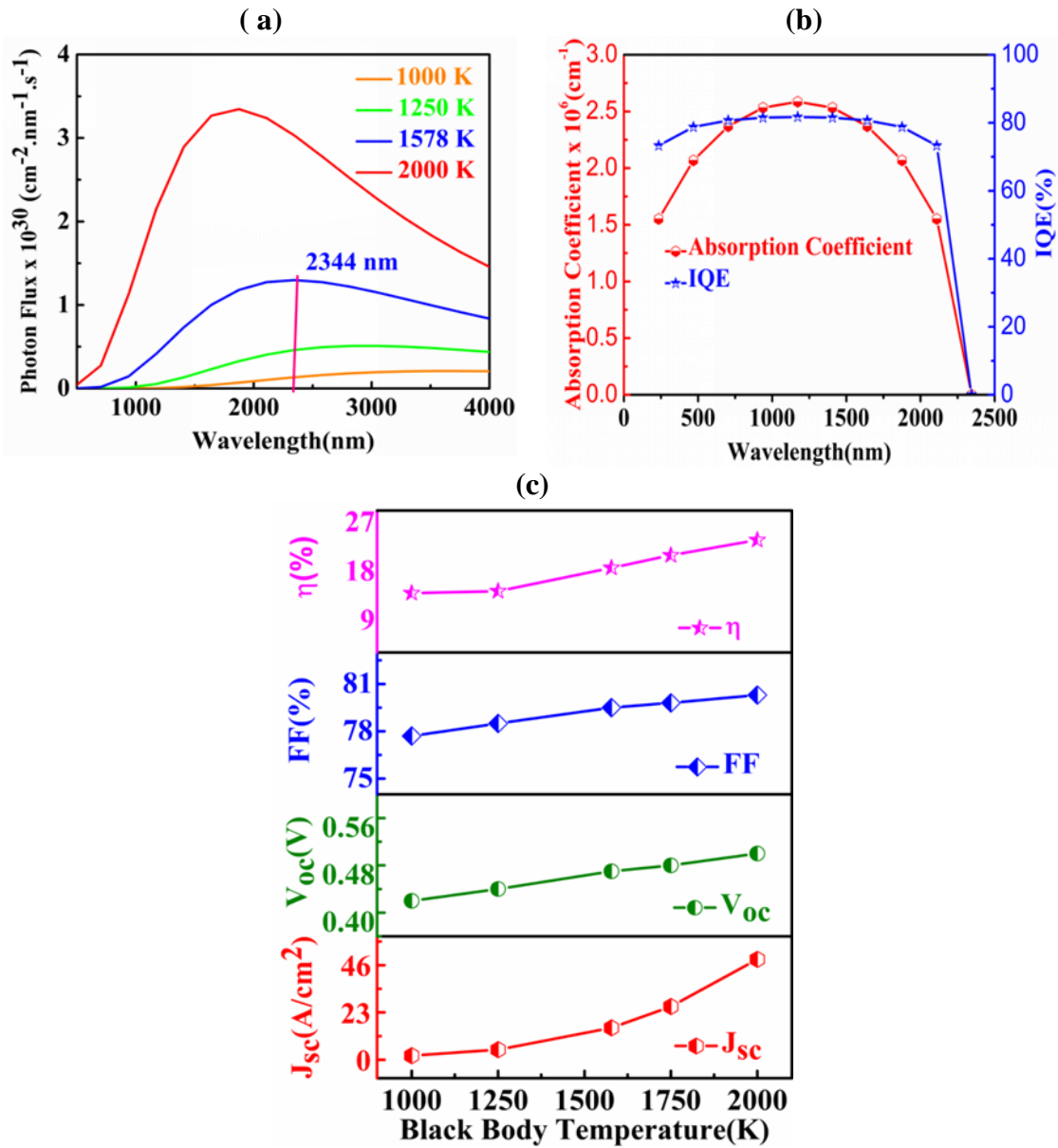

**Figure 11:** (a) Variations in photon flux relative to the spectral length at varying temperatures, (b) Adjustments in the coefficient of the absorption and the IQE with respect to wavelength, and (c) Changes in output parameters for the TPV system in relation to black body temperature.

Figure 11(b) illustrates the variations in the coefficient of the absorption and the IQE over a wavelength spectrum spanning from 234 nm to the cutoff spectral response of 2344 nm. Initially, the absorption coefficient rises from $1.5\times10^6$ cm$^{-1}$ to $2.5\times10^6$ cm$^{-1}$, peaking around 1250 nm. Subsequently, it marginally decreases, ultimately reaching zero at the cutoff wavelength, indicating no further absorption beyond this point. The behavior of the absorption coefficient concerning wavelength is governed by the formula αhv=A(hv-Eg)$^P$, where A represents a constant, and P denotes an exponent [37]. Furthermore, the initial IQE point stands at approximately 75%, starting at 230 nm. It then climbs to a peak of 80% at 1250 nm, followed by a gradual decline to nearly 75% as it nears the cutoff wavelength. Beyond the cutoff wavelength, there is an abrupt drop to 0%, which enhances the dark current, presenting a disadvantageous scenario.

Figure 11(c) illustrates the significant influence of the $T_{BB}$ on the output parameters of the InGaAsSb cell. While two parameters show substantial improvement, the other two demonstrate a more moderate enhancement as the black-body temperature increases from 1000 K to 2000 K. Within the studied range of black body temperature, $J_{SC}$ experiences a remarkable change, ranging between almost 0.5 A/cm$^2$ and 46.5 A/cm$^2$. This occurs due to the rise in photon flow as $T_{BB}$ increases, leading to a significant augmentation in $J_{SC}$ within the device [1]. $V_{OC}$ follows a moderate upward path, starting from 0.4 V and ending at 0.5 V. The carrier collection increases with $T_{BB}$, resulting in the band filling effect. As a consequence of the enhanced maximum power, the fill factor experiences a slight improvement, increasing from 78% to 81%. The elevation of these three variables propels the outcome of the TPV cell to rise from 10% to approximately 24%.

### 3.3.4 SRH and the intrinsic carrier impact on the cell operation

The TPV parameters are highly sensitive to the SRH recombination and the intrinsic carrier, $n_i$ which is shown in the Figure 12. The figures show domination of SRH lifetime is towards the x axis, whereas, the domination of intrinsic carrier is towards the y axis.

With this two variables, the reaction of $J_{SC}$ is recorded in Figure 12(a). Only a slight variation of the $J_{SC}$ is observed in this figure which ranging between 15.3 A/cm$^2$ to 15.8 A/cm$^2$ when the SRH carrier lifetime spanning from 0.1 μs to 1.5 μs, and $n_i$ spanning from $4.5\times10^{11}$ cm$^{-3}$ to $4.5\times10^{15}$ cm$^{-3}$. At $4.5\times10^{13}$ cm$^{-3}$, when the SRH lifetime rises from 0.1 μs to 1.5 μs, the $J_{SC}$ remains constant at 15.55 A/cm$^2$. At the same time, when SRH stays at 1 μs, the variation of $n_i$ at the observed boundary causes the $J_{SC}$ to fall from 15.7 A/cm$^2$ to 15.4 A/cm$^2$. Those observations indicate, the $J_{SC}$ is slightly related with those variables.

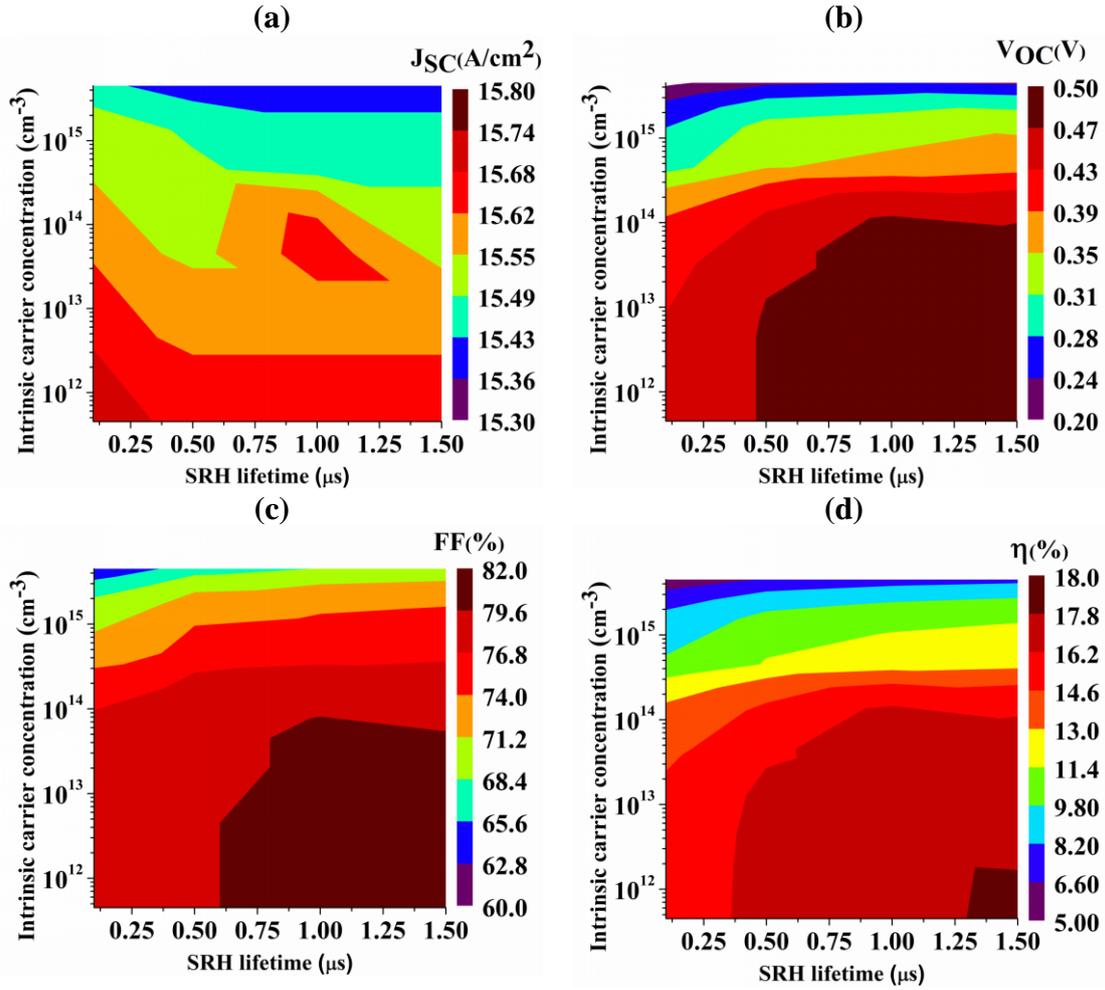

**Figure 12:** The fluctuations in output parameters: (a) $J_{SC}$, (b) $V_{OC}$, (c) FF, and (d) η concerning the intrinsic carrier and the SRH lifetime.

The Figure 12(b) attests that the $V_{OC}$ is slightly sensitive to the SRH recombination, whereas, it is highly reactive with the $n_i$. At $4.5 \times 10^{13}$ cm$^{-3}$ the rise of SRH lifetime from 0.1 μs to 1.5 μs, causes the $V_{OC}$ to reach 0.47 V from 0.5 V. However, at a constant SRH lifetime of 1 μs, when the $n_i$ reaches the edge of the observed boundary the $V_{OC}$ falls from 0.5 V to 0.26 V. The amelioration of the dark current with the augmentation of $n_i$ takes the responsibilities to the huge loss of the $V_{OC}$ [1]. It's not surprising that, the FF emulates the behavior of the $V_{OC}$, which can be examined in Figure 12(c). The span of FF is from 69% to 81%. The domination of the $V_{OC}$ forces the efficiency to follow its path which is showed in 12(d). At the constant $n_i$ of $4.5 \times 10^{13}$ cm$^{-3}$, the variation of the SRH lifetime pulls the efficiency from 14.2 % to 16.9 % as it maintains a positive relation with the carrier lifetime which works for the betterment of the efficiency [1]. Inversely, at constant SRH lifetime of 1 μs, when the $n_i$ approaches the top from the bottom, the efficiency drastically falls to 6.6% from about 17% as it's related to the Voc.

## 3.3.5 Radiative and Auger recombination impact on the output parameters

Figure 13 demonstrates the changes in output parameters concerning variations in radiative recombination ($B_{opt}$) along the x-axis and shifts in Auger recombination (**A**) along the y-axis. The assessed interval for the $B_{opt}$ and **A** ranges from $10^{-16}$ cm$^3$s$^{-1}$ to $10^{-12}$ cm$^3$s$^{-1}$ and from $10^{-33}$ cm$^6$s$^{-1}$ to $10^{-29}$ cm$^6$s$^{-1}$, respectively. At constant Auger recombination coefficients of $10^{-31}$ cm$^6$s$^{-1}$, when the $B_{opt}$ varies from $10^{-16}$ cm$^3$s$^{-1}$ to $10^{-12}$ cm$^3$s$^{-1}$ the $J_{SC}$ stays almost same at 15.55 A/cm$^2$, however just crossing $10^{-12}$ cm$^3$s$^{-1}$ it rises up to 15.6 A/cm$^2$ which is presented in Figure 13(a). Inversely, at a stable value of the radiative recombination varies of $10^{-14}$ cm$^3$s$^{-1}$, when the Auger recombination coefficients varies from the bottom to the top, the $J_{SC}$ rises from 15.53 A/cm$^2$ to 15.63 A/cm$^2$.

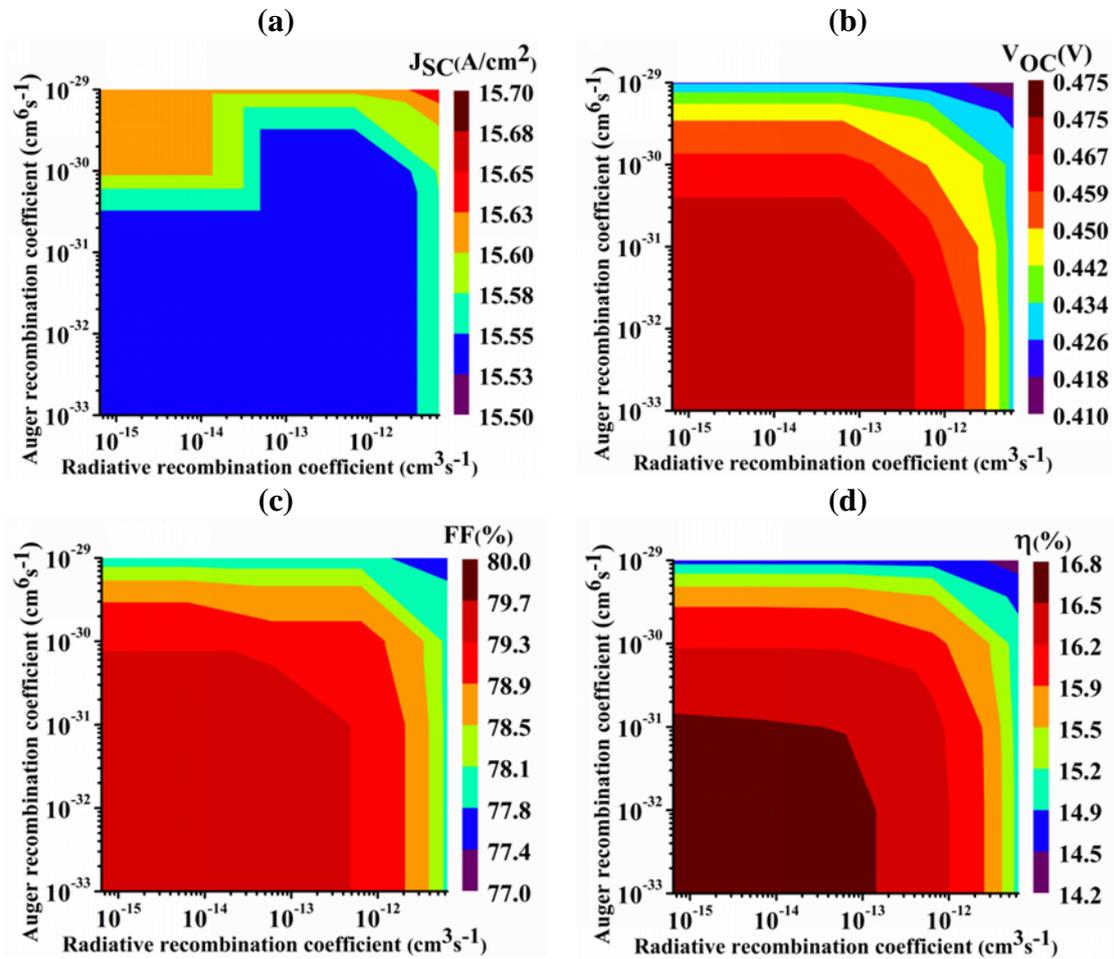

**Figure 13:** The fluctuations in output parameters: (a) $J_{SC}$, (b) $V_{OC}$, (c) FF, (d) η concerning the radiative and the Auger recombination coefficient.

Figure 13(b) illustrates that while $V_{OC}$ shows very sensitivity to the $B_{opt}$, and **A**. With $B_{opt}$ increasing from $10^{-16}$ cm3s-1 to $10^{-13}$ cm$^3$s$^{-1}$ at a constant **A** of $10^{-31}$ cm$^6$s$^{-1}$, $V_{OC}$ slightly falls from 0.475 V to 0.47 V. Additionally, with $B_{opt}$ increasing from $10^{-13}$ cm$^3$s$^{-1}$ to $10^{-12}$ cm$^3$s$^{-1}$ the $V_{OC}$ moderately falls from 0.47 V to 0.42 V. Conversely, at a constant $B_{opt}$ of 6.4X10$^{-14}$ cm3s$^{-1}$, as **A** approaches $10^{-31}$ cm$^6$s$^{-1}$ from $10^{-33}$ cm$^6$s$^{-1}$, $V_{OC}$ slightly declines from 0.475 V to 0.47 V. Nevertheless after crossing this boundary it plummets surprisingly from 0.47 V to 0.42 V at $10^{-29}$ cm$^6$s$^{-1}$. The FF behaves similarly to $V_{OC}$, as shown in Figure 13(c), ranging from 77% to 80%. The dominance of $V_{OC}$ dictates the efficiency trend which is depicted in Figure 13(d). With a constant **A** of $10^{-31}$ cm$^6$s$^{-1}$, changes in $B_{opt}$ alter the efficiency from 16.5% to 14.9%. Radiative recombination is associated with the lifetime of minority charge carriers using the identical equation that governs the Auger recombination coefficient [38]. Conversely, at a constant $B_{opt}$ of $10^{-14}$ cm3s$^{-1}$, a significant efficiency drop from around 17% to 14.5% occurs as A increases. A higher magnitude of Auger recombination decreases the lifespan of the minority carriers charge, resulting in a decline in all the parameters [38].

**3.3.6 Cell temperature impact on the device performance**

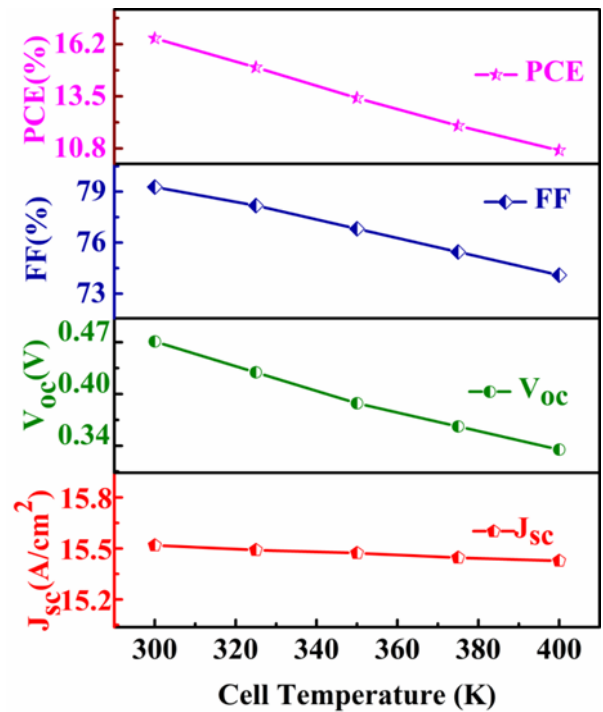

**Figure 14.** Alterations in output parameters InGaAsSb TPV device related to cell temperature.

The Figure 14 tells how the resultant parameters are influenced by the cell temperature within the range of 300 K to 400 K. It's evident from the figure that the $J_{SC}$ stays mostly stable despite changes in cell temperature. Consequently, a linear decrease in $V_{OC}$ is observed within this range as the $V_{OC}$ is significantly affected by the cell temperature, ranging from about 0.47 V to 0.33 V. The rise in the temperature of the cell can escalate the dark current, consequently reducing the $V_{OC}$. The FF of the TPV device follows a similar trend as $V_{OC}$ due to their empirical relationship. The decline in $V_{OC}$ demonstrates its impact on efficiency, dropping from 16.5% at 300 K to 10.7% at 400 K. Hence, maintaining an increased proportion between the $V_{OC}$ and the energy-gap is essential to reduce the cell's susceptibility to fluctuations in the temperature of the TPV cell [38].

### 3.3.7 Optimization of InGaAsSb based homo and hetero-junction

Figure 15 presents the J-V curve of InGaAsSb based thermophotovoltaic cell, where the homo-junction structured p-InGaAsSb/n-InGaAsSb and the hetero-junction structured p-Ge/n-InGaAsSb. The InGaAsSb single homo-junction TPV cell achieves an efficiency of 12.6% with $J_{SC}$ at 15.24 A/cm$^2$, $V_{OC}$ at 0.38 V, and FF at 76.3%. Introducing Ge as a window layer significantly boosts efficiency. The Hetero TPV cell, for instance, achieves an efficiency of approximately 16.5% with a $J_{SC}$ at 15.53 A/cm$^2$, $V_{OC}$ at 0.47 V, and FF at 79.5%. The escalation of the built-in potential and hence the reduction of dark current enhance the open circuit voltage which in turn raises the performance of the single hetero-junction thermophovoltaic cell [39].

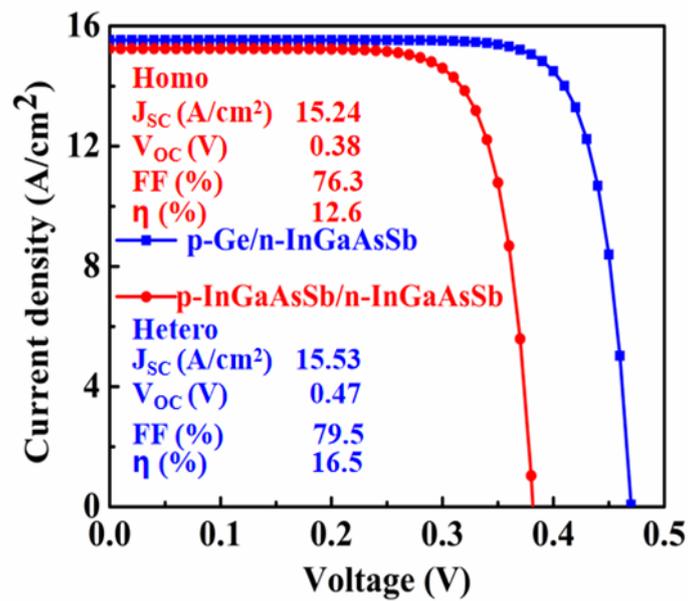

**Figure 15.** J-V characteristic comparison of InGaAsSb-based homo and hetero-junction.

### 4. Conclusion

This research underscores the promising potential of InGaAsSb as a material applicable in Thermophotovoltaic (TPV) systems. By employing Machine Learning calculations, the study delves into the variations of inputs within conventional TPV cells, providing insights into their inner workings and how outputs depend on these inputs. The aim is to identify the most effective hetero-junction combination among different materials. Ge emerges as the optimal window layer, paired with InGaAsSb as the best base layer. Utilizing a single-hetero-junction p-Ge/n-InGaAsSb TPV cell with a p-n structure, and leveraging machine learning techniques, the work assesses the significance of various parameters in the device transport model, thus enhancing comprehension and enabling design optimization. The optimized TPV cell demonstrates notable enhancements, achieving an efficiency of 16.50%, with key parameters such as $J_{SC}$, $V_{OC}$, and FF are about 15.53 A/cm$^2$, 0.47 V, and 79.5%, respectively. These findings reveal the importance of integrating computational and experimental methods to advance TPV materials and devices.

***Corresponding author:**
E-mail: jak_apee@ru.ac.bd (Jaker Hossain).

**Notes:** The authors declare no competing financial interest.